\documentclass[aps,pre,preprintnumbers,floats,twocolumn,superscriptaddress,nofootinbib,longbibliography]{revtex4}
\usepackage{graphicx}
\usepackage{dcolumn}
\usepackage{bm}
\usepackage[english]{babel}
\usepackage[latin1]{inputenc}
\usepackage{graphicx}
\usepackage{epstopdf}
\usepackage{amsfonts}
\usepackage{comment}
\usepackage{url}
\usepackage{color}
\usepackage{epsfig}
\usepackage{mathrsfs}
\usepackage{amsmath,amssymb}
\usepackage{subfigure}

\usepackage{framed}
\usepackage[svgnames]{xcolor}
\definecolor{shadecolor}{named}{LightGrey}

\newcommand{\lto}[1]{\longrightarrow#1}

\renewcommand{\(}{\left(}
\renewcommand{\)}{\right)}
\renewcommand{\[}{\left[}
\renewcommand{\]}{\right]}

\newcommand{\rev}[1]{{\color{blue}\textbf{Manlio:} #1}}

\DeclareMathOperator*{\argmin}{arg\,min}

\begin{document}

\graphicspath{{figure/}}
\selectlanguage{english}


\title{Centrality in Interconnected Multilayer Networks}

\author{Manlio De Domenico}
 \affiliation{Departament d'Enginyeria Inform\`atica i Matem\`atiques, Universitat Rovira i Virgili, 43007 Tarragona, Spain}

\author{Albert Sol\'e-Ribalta}
 \affiliation{Departament d'Enginyeria Inform\`atica i Matem\`atiques, Universitat Rovira i Virgili, 43007 Tarragona, Spain}

\author{Elisa Omodei}
 \affiliation{ISC-PIF, LaTTiCe-CNRS, \'Ecole Normale Sup\'erieure, Paris, France}

\author{Sergio G\'omez}
 \affiliation{Departament d'Enginyeria Inform\`atica i Matem\`atiques, Universitat Rovira i Virgili, 43007 Tarragona, Spain}

\author{Alex Arenas}
 \affiliation{Departament d'Enginyeria Inform\`atica i Matem\`atiques, Universitat Rovira i Virgili, 43007 Tarragona, Spain}


\begin{abstract}
Real-world complex systems exhibit multiple levels of relationships. In many cases, they require to be modeled by interconnected multilayer networks, characterizing interactions on several levels simultaneously. It is of crucial importance in many fields, from economics to biology, from urban planning to social sciences, to identify the most (or the less) influent nodes in a network. However, defining the centrality of actors in an interconnected structure is not trivial. 

In this paper, we capitalize on the tensorial formalism, recently proposed to characterize and investigate this kind of complex topologies, to show how several centrality measures -- well-known in the case of standard (``monoplex'') networks -- can be extended naturally to the realm of interconnected multiplexes. We consider diagnostics widely used in different fields, e.g., computer science, biology, communication and social sciences, to cite only some of them. We show, both theoretically and numerically, that using the weighted monoplex obtained by aggregating the multilayer network leads, in general, to relevant differences in ranking the nodes by their importance.
\end{abstract}

\maketitle

\flushbottom




\section{Introduction}

It is common practice in many studies involving networks to assume that nodes are connected to each other by a single type of static edge that encapsulates all connections between them. In a myriad of applications this assumption oversimplifies the complexity of the network, leading in some cases to misleading results. A representative example is provided by temporal networks, where neglecting time-dependence washes out the memory of sequences of human contacts in transmission of diseases \cite{holme2012}. Similarly, neglecting the existence of multiple relationships between actors might alter the topology of and the dynamics on the top of networks, leading to overestimation (or underestimation) of crucial properties of nodes, as their centrality \cite{freeman1979centrality,jeong2001lethality,guimera2002optimal,guimera2005worldwide,barthelemy2004betweenness,nicosia2012controlling} with respect to specific criteria (e.g., communicability, influence, \emph{etc}). In the specific case of multilayer networks, understanding the centrality of nodes is not trivial.

Before going into the details of the present study, it is important to discuss the difference between the topological structure which represents the core of this study, namely interconnected multilayer networks \cite{mucha2010community,gomez2013diffusion,radicchi2013abrupt,dedomenico2013mathematical,sole2013spectral,granell2013interplay,garcia2013dimensionality}, and other multilayer structures which have been named \emph{multiplexes} in the past and have been the subject of recent studies \cite{lee2012,nicosia2013growing,Bianconi2013Statistical,battiston2013metrics,Sola2013Centrality}. Note that interconnected multilayer networks are not simply a special case of or equivalent to interdependent networks \cite{gao2011networks}: in multilayer systems, many or even all of the nodes have a counterpart in each layer, so one can associate a vector of states to each node. This feature has no counterpart in interdependent networks, which were conceived as interconnected communities within a single, larger network \cite{buldyrev2010,dickison2012epidemics}.

Historically, the term \emph{multiplex} has been adopted to indicate the presence of more than one relationship between the same actors of a social network \cite{padgett1993robust}. This type of network is well understood in terms of ``coloring'' (or labeling) the edges corresponding to interactions of different nature. For instance, the same individual might have connections with other individuals based on financial interests (e.g., color red) and connections with the same or different individuals based on friendship (e.g., color blue). This type of network is represented by a \emph{non-interconnected multiplex}.

\begin{figure}[!t]
	\centering
	  \includegraphics[width=8cm]{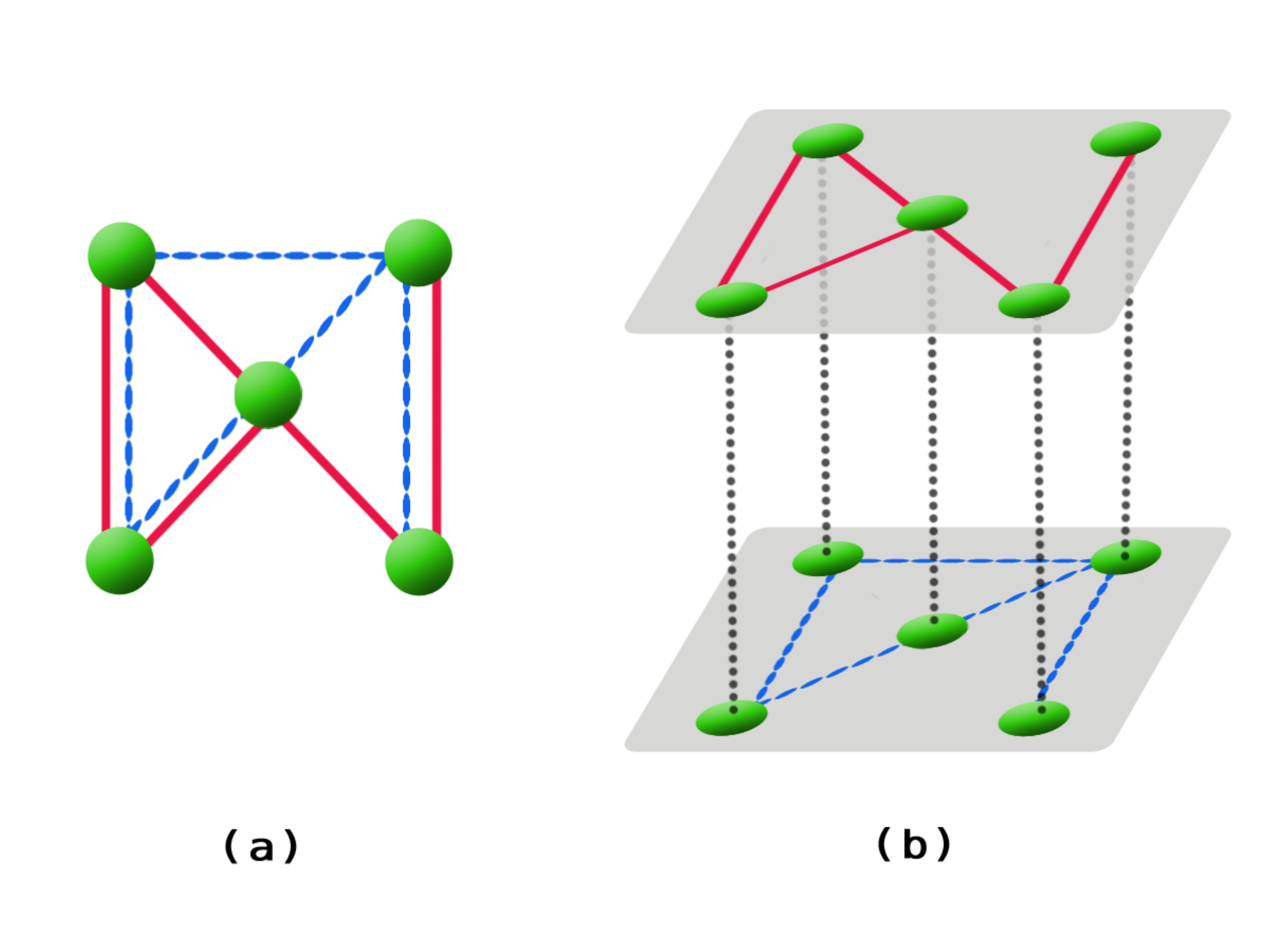}
	\caption{a) Edge-colored graph (i.e., multiplex) representing two different types of interactions (solid and dashed edges) between 5 actors. b) An interconnected multiplex representing the same actors exhibiting the same relationships but on different levels which are separated by a cost (dotted vertical lines) to move from one layer to the other.}
        \label{fig:mplex-vs-edgecolor}
\end{figure}

Conversely, in other real-world systems, like the transportation network of a city, the same geographical position can be part, for instance, of the network of subway or the network of bus routes, simultaneously. In this specific case, an edge-colored graph would not capture the full structure of the network, because it is missing information about the cost to \emph{move} from the subway network to the bus route. This cost can be economic or might account for the time required to physically commute between the two layers. Therefore, the interconnected multilayer topology presented in this section provides a better representation of the system. In Fig.\,\ref{fig:mplex-vs-edgecolor} is shown an illustration of an edge-colored graph (Fig.\,\ref{fig:mplex-vs-edgecolor}a) and an interconnected multiplex (Fig.\,\ref{fig:mplex-vs-edgecolor}b). It is evident that a simple projection of the latter -- mathematically equivalent to sum up the corresponding adjacency matrices -- would provide a network where the information about the colors is lost. On the other hand, an edge-colored graph can not account for interconnections, keeping unreconcilable the two structures in Fig.\,\ref{fig:mplex-vs-edgecolor} which should be used to represent very different networked systems.

For further details about the classification of such multilayer networks we refer to \cite{kivela2013multilayer} and references therein.

In this paper, we extend widely adopted measures of centrality to interconnected multilayer networks. We consider diagnostics based either only on topology or on both topology and dynamics on the top of the network. In the specific case of descriptors based on dynamics, we validate our theory against detailed simulations. In fact, we show analytically and numerically that calculation of centrality of nodes in multilayer networks can not prescind from considering the existence of the interconnections between different layers. The use of the weighted monoplex obtained by aggregating the interconnected multilayer network might lead, in general, to relevant differences in the ranking of the nodes.

The remainder of this paper is organized as follows. In Sec.\,\ref{sec:tensors} we briefly describe the tensorial notation, defined in \cite{dedomenico2013mathematical}, adopted in the rest of the paper. In Sec.\,\ref{sec:centrality} we capitalize on this notation to extend many well-known centrality descriptors, defined in the case of monoplex networks, to the realm of interconnected multilayer networks. Finally, we discuss our findings in Sec.\,\ref{sec:conclusion}.


\section{Tensorial notation}\label{sec:tensors}

Edge-colored graphs can be represented by a set of adjacency matrices \cite{nicosia2013growing,Bianconi2013Statistical,battiston2013metrics}. However, standard matrices, used to represent networks, are inherently limited in the complexity of the relationships that they can capture, i.e., they do not represent a suitable framework in the case of interconnected multiplexes. This is the case of increasingly complicated types of relationships -- that can also change in time -- between nodes. Such a level of complexity can be characterized by considering tensors and algebras of higher order \cite{dedomenico2013mathematical}. 

A great advantage of tensor formalism also relies on its compactness. An adjacency tensor can be written using a more compact notation that is very useful for the generalization to \emph{multilayer} networks. In this notation, a row vector $\mathbf{a}\in\mathbb{R}^{N}$ is given by a covariant vector $a_{\alpha}$ ($\alpha=1,2,\ldots,N$), and the corresponding contravariant vector $a^{\alpha}$ (i.e., its dual vector) is a column vector in Euclidean space. A canonical vector is assigned to each node and the corresponding interconnected multi-layer network is represented by a rank-4 adjacency tensor.

However, in the majority of applications, it is not necessary to perform calculations using canonical vectors and tensors explicitly. Consequently, a classical single-layer network represented by a rank-2 mixed adjacency tensor $W^{\alpha}_{\beta}$ \cite{dedomenico2013mathematical} can be simply indicated by $W^{i}_{j}$, where the ``abuse of notation'' consists in interpreting the indices $i$ and $j$ as nodes and $W^{i}_{j}$ would indicate intensity of the relationship between them. Hence, $W^{i}_{j}$ represents the well-known adjacency matrix of a graph and the classical notation for the weight $w_{ij}$ of the link between $i$ and $j$ corresponds to $W^{i}_{j}$. The ``abuse of notation'' also consists in treating $W^{i}_{j}$ as a rank-2 tensor, although it explicitly indicates the entry of a matrix, while keeping the algebraic rules governing covariant and contravariant tensors. This ``abuse of notation'' dramatically reduces the complexity of some tensorial equations, although it is worth remarking that it should be used only when calculations do not involve canonical tensors explicitly.

To distinguish simple networks from the more complicated situations (e.g., interconnected multiplex networks) that we use in this paper, we will use the term \emph{monoplex networks} to describe such standard networks, which are time-independent and possess only a single type of edge that connects its nodes.

In general, there might be several types of relationships between pairs of nodes and a more general system represented as a multilayer object -- in which each type of relationship is encompassed in a single \emph{layer} $\alpha$ ($\alpha=1,2,\ldots,L$) of a system -- is required. Note that $\alpha$ has no more the same meaning of the index in the adjacency tensor discussed above. To avoid confusion, in the following we refer to nodes with Latin letters and to layers with Greek letters, allowing us to distinguish indices that correspond to nodes from those that correspond to layers in tensorial equations.

We use an \emph{intra-layer adjacency tensor} for the $2^{\text{nd}}$-order tensor $W^{i}_{j}(\alpha)$ that indicates the relationships between nodes within the \emph{same} layer $\alpha$. We take into account the possibility that a node $i$ from layer $\alpha$ can be connected to any other node $j$ in any other layer $\beta$. To encode information about relationships that incorporate multiple layers, we introduce the $2^{\text{nd}}$-order \emph{inter-layer adjacency tensor} $C^{i}_{j}(\alpha\beta)$. Note that $C^{i}_{j}(\alpha\alpha)=W^{i}_{j}(\alpha)$.

It has been shown that the mathematical object accounting for the whole interconnected multilayer structure is given by a $4^{\text{th}}$-order (i.e., rank-4) \emph{multilayer adjacency tensor} $M^{i\alpha}_{j\beta}$. This tensor might be simply thought as a higher-order matrix with four indices. It is the direct generalization of the adjacency matrix in the case of monoplexes, encoding the intensity of the relationship (which may not be symmetric) between a node $i$ in layer $\alpha$ and a node $j$ in layer $\beta$ \cite{dedomenico2013mathematical}. This object is very general and can be used to represent structures where an actor is present in some layers but not in all of them. This is the case, for instance when considering a network of online social relationships, of an individual with an account on Facebook but not on Twitter. The algebra still holds for these situations without any formal modification. In fact, one simply introduces ``empty nodes'' and assigns the value $0$ to the associated edges, although the calculations of network diagnostics should carefully account for the presence of such nodes (for instance, for a proper normalization) \cite{dedomenico2013mathematical}.

Often, to reduce the notational complexity in the tensorial equations, the Einstein summation convention is adopted. 
It is applied to repeated indices in operations that involve tensors. For example, we use this convention in the left-hand sides of the following equations:
\begin{align}
	A^{i}_{i} = \sum_{i=1}^{N}A^{i}_{i}\,,\quad
	A^{i}_{j}B^{j}_{i} =\sum_{i=1}^{N}\sum_{j=1}^{N}A^{i}_{j}B^{i}_{j}\,,\nonumber\\
	A^{i\alpha}_{j\beta}B^{k\beta}_{i\gamma} =\sum_{i=1}^{N}\sum_{\beta=1}^{L}A^{i\alpha}_{j\beta}B^{k\beta}_{i\gamma}\,,\nonumber
\end{align}
whose right-hand sides include the summation signs explicitly.  It is straightforward to use this convention for the product of any number of tensors of any order. In the following, we will use the $t$-th power of rank-4 tensors, defined by multiple tensor multiplications:
\begin{equation}
(A^t)^{i\alpha}_{j\beta} = (A)^{i\alpha}_{j_1\beta_1}(A)^{j_1\beta_1}_{j_2\beta_2}\dots(A)^{j_{t-1}\beta_{t-1}}_{j\beta}
\end{equation}

Repeated indices, such that one index is a subscript and the other is a superscript, is equivalent to perform a tensorial operation known as a \emph{contraction}. Moreover, one should be very careful in performing tensorial calculations. For instance, using traditional notation the product $a^{i}b^{j}$ would be a number, i.e., the product of the components of two vectors. However, in our formulation, the same calculation denotes a Kronecker product between two vectors, resulting in a rank-2 tensor, i.e., a matrix.

An interesting network that can be derived from the interconnected structure is the aggregated network, where the edges between two actors are summed up across all layers. The superposition of the different layers is equivalent to summing up the adjacency tensor of each layer. The corresponding aggregated network $G^{i}_{j}$ is a monoplex and is obtained by contracting the layer indices of the multilayer adjacency tensor, i.e., $G^{i}_{j}=M^{i\alpha}_{j\alpha}$. This aggregation loses the information about inter-layer connections. If such an information is important for the application of interest, then the tensor should be contracted with the 1-tensor $u^{\beta}_{\alpha}$ (the rank-2 tensor with all components equal to 1), i.e., $\bar{G}^{i}_{j}=M^{i\alpha}_{j\beta}u^{\beta}_{\alpha}$.

This formalism is extremely useful to put in evidence how topological descriptors of interconnected networks differ from the ones corresponding to their aggregated graphs \cite{dedomenico2013mathematical,cozzo2013cc}. Moreover, it is particularly suitable to perform compact calculations. 

As a representative example, let us consider the number of paths of length 2 from a node in a certain layer to any other node in any other layer of the system. Taking advantage of the extended algebra, it is straightforward to show that the resulting rank-4 tensor accounting for such paths is given by $H^{i\alpha}_{j\beta}=M^{i\alpha}_{k\gamma}M^{k\gamma}_{j\beta}$. If only the number of paths between any pair of nodes is required, regardless of the layer, then the corresponding rank-2 tensor of paths is simply obtained by contracting with the 1-tensor $u^{\beta}_{\alpha}$, i.e., $X^{i}_{j}=H^{i\alpha}_{j\beta}u^{\beta}_{\alpha}$. Conversely, in the case of the aggregate, we first contract the multilayer adjacency tensor to obtain the aggregation $J^{i}_{j}=M^{i\alpha}_{j\beta}u^{\beta}_{\alpha}$, where inter-layer connections are included as self-loops, and then square the resulting tensor to obtain $Y^{i}_{j}=J^{i}_{k}J^{k}_{j}$. Of course, a similar argument can be used to calculate the number of longer paths. From these tensorial equations it is evident that the aggregated graph can not be considered, in general, a good proxy of the interconnected topology.

Summarizing, the tensorial formulation provides a suitable framework for several real-world networked systems, from transportation networks to social ones. It is also worth noting that special cases of multilayer adjacency tensors are time-dependent (i.e., ``temporal") networks \cite{dedomenico2013mathematical,kivela2013multilayer}.
More specifically, in the case of social sciences the multilayer adjacency tensor can be used, for instance, to model the structural changes of a social network over time, or to define the topology of actors involved in several different levels of relationships and for whom it is indispensable to define an inter-connection between such levels. For these networked systems, it is desirable to adopt descriptors (e.g., clustering coefficient, modularity, \emph{etc}) that are the natural extension of their well-known counterparts in monoplex networks.


\section{Centrality in Interconnected Networks}\label{sec:centrality}

In this section, we focus on the definition of node centrality in a multilayer network. We obtain these properties using algebraic operations involving the multilayer adjacency tensor, canonical vectors, and canonical tensors, achieving the natural extension of the concept of centrality in single-layer networks. We refer to \cite{dedomenico2013mathematical} for other multilayer network diagnostics.

In practical applications one is often interested in assigning a global measure of importance to each node, aggregating the information obtained from the different layers. A naive choice could be to combine the centrality of the nodes -- obtained from the different layers separately -- according to some heuristic choice. This is a viable solution when there is no interconnection between layers, i.e., in the case of edge-colored graphs \cite{Sola2013Centrality,pagerank2013}. However, the main drawback of applying this approach to interconnected multilayer networks is that the measure will depend on the choice of the heuristics and might not evaluate the real importance of nodes. Conversely, our approach capitalizes on the tensorial formulation of interconnected multilayer networks and accounts for the higher level of complexity of such systems without relying on external assumptions and naturally extending the well-known centrality measures adopted for several decades in the case of monoplexes.

\subsection{Centrality based on dynamical properties}

\emph{Random walk occupation centrality.} A random walk is the simplest dynamical process that can occur on a monoplex network, and random walks can be used to approximate other types of diffusion \cite{chung1997,Newman2010Book}. Random walks on monoplex networks \cite{chung1997,noh2004random,Newman2010Book} have attracted considerable interest because they are both important and easy to interpret. They have yielded important insights on a huge variety of applications and can be studied analytically. For example, random walks have been used to rank Web pages \cite{pagerank1998} and sports teams \cite{callaghan2007}, optimize searches \cite{viswanathan1999optimizing}, investigate the efficiency of network navigation
\cite{yang2005exploring,da2007exploring}, characterize cyclic structures in networks \cite{rozenfeld2005statistics}, and coarse-grain networks to illuminate meso-scale features such as community structure \cite{gfeller2007spectral,rosvall2007information,lambiotte2008}. Another interesting application of random walks is to calculate the centrality of actors in complex networks when they have not knowledge of the full topology but only local information is available. In such cases, centrality descriptors based on shortest-paths, e.g., betweenness and closeness centrality, should be substituted by centrality notions based on random walks \cite{noh2004random,newman2005measure}. We extend these measures to interconnected multilayer networks further in the text.

\begin{figure}[!t]
	\centering
	  \includegraphics[width=8cm]{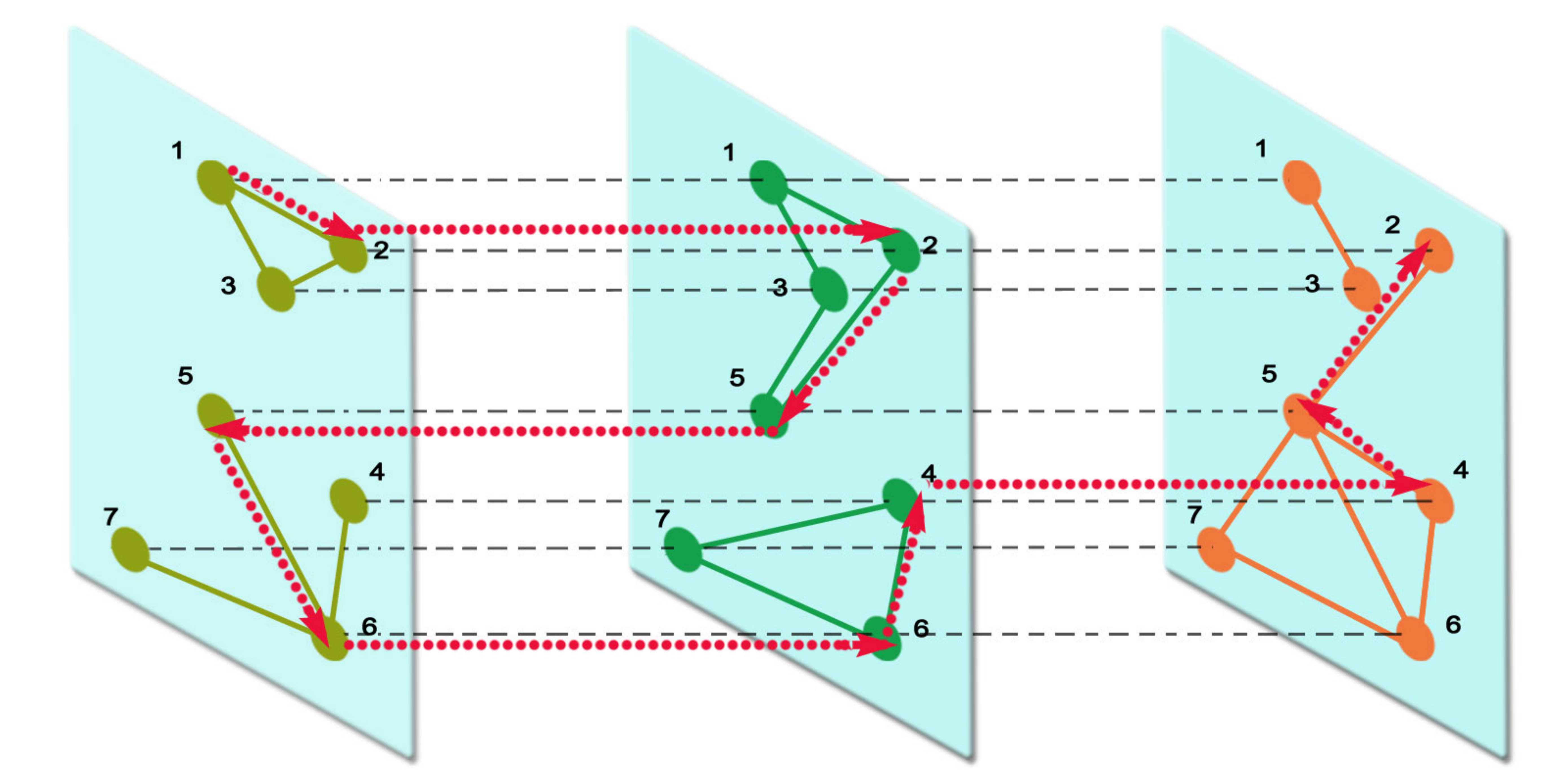}
	\caption{Schematic of a random walk (dotted trajectories) in a multiplex network. A walker can jump between nodes within the same layer, or it might switch to another layer. This illustration evinces how multiplexity allows a random walker to move between nodes that belong to different (disconnected) components on a given layer.}
        \label{fig:mplex-rw}
\end{figure}

In this paper, we consider a discrete-time random walk. As we illustrate in Fig.~\ref{fig:mplex-rw}, a random walk on a multilayer network induces nontrivial effects because the presence of inter-layer connections affects its navigation of a networked system \cite{dedomenico2013random}. Let $T^{i\alpha}_{j\beta}$ denote the tensor of transition probabilities for jumping between pairs of nodes and switching between pairs of layers, and let $p_{i\alpha}(t)$ be the time-dependent tensor that gives the probability to find a walker at a particular node in a particular layer. Hence, the covariant master equation that governs the discrete-time evolution of the probability from time $t$ to time $t+1$ is $p_{j\beta}(t+1)=T^{i\alpha}_{j\beta}p_{i\alpha}(t)$.  

The steady-state solution of this equation, i.e., for $t\lto\infty$, is given by $\Pi_{i\alpha}$, quantifying the probability to find a walker in the node $i$ of layer $\alpha$. In the case of monoplexes, the steady-state solution can be obtained by solving the eigenvalue problem for the rank-2 transition tensor and calculating the leading eigenvector corresponding to the unitary eigenvalue. Similarly, in the case of multilayer networks, the solution can be obtained by calculating the leading \emph{eigentensor}, solution of the higher-order eigenvalue problem
\begin{eqnarray}
T^{i\alpha}_{j\beta}\Pi_{i\alpha}=\lambda\Pi_{j\beta}.
\end{eqnarray}
We refer to Appendix\,\ref{sec:eigentensor} for the mathematical details to solve this problem.

The probability $\Pi_{j\beta}$, that we define \emph{random walk occupation centrality}, accounts for the full interconnected structure of the multilayer network. Although different exploration strategies can be adopted to walk in a multilayer network \cite{dedomenico2013random}, here we focus on the natural extension of well-known random walks in monoplex networks \cite{noh2004random}. In this process, the walker in node $i$ and layer $\alpha$ might jump to one of its neighbors $j\neq i$ -- within the same layer -- with uniform probability, or might switch to its counterpart $i$ in a different interconnected layer $\beta\neq\alpha$. It is worth remarking that the inter-layer connection is treated as an edge that can be chosen randomly among all edges traversing the node.

In the more general case of weighted networks, the jumping probability is proportional to the weight of the edges. Let us indicate with $s_{i\alpha}$ the strength of node $i$ in layer $\alpha$, including the inter-layer connections. The multi-strength vector, whose components indicate the strength of each node accounting for the full multilayer structure, is given by summing up its strengths across all layers, i.e., by $S_{i}=s_{i\alpha}u^{\alpha}$, where $u^{\alpha}$ is the 1-vector, namely a vector with all components equal to 1. We indicate with $D^{i\alpha}_{j\beta}$ the strength tensor whose entries are all zeros, except for $i=j$ and $\alpha=\beta$ where the entries are given by $s_{i\alpha}$. This tensor represents the multilayer extension of the well-known diagonal strength matrix in the case of monoplexes. Therefore, the transition tensor is given by $T^{i\alpha}_{j\beta}=M^{k\gamma}_{j\beta}\tilde{D}^{i\alpha}_{k\gamma}$, where $\tilde{D}^{i\alpha}_{j\beta}$ is the tensor whose entries are the inverse\footnote{It is worth remarking that, in general, this is different from the inverse of a tensor $A^{i\alpha}_{j\beta}$, that is defined as the tensor $B^{i\alpha}_{j\beta}$ such that $A^{i\alpha}_{k\gamma}B_{j\beta}^{k\gamma}=\delta^{i\alpha}_{j\beta}$, where $\delta^{i\alpha}_{j\beta}=\delta^{i}_{j}\delta^{\alpha}_{\beta}$.} of the non-zero entries of the strength tensor. For this classical random walk, it can be easily shown that $\Pi_{i\alpha}\propto s_{i\alpha}$ \cite{dedomenico2013random}.

This centrality, as others in the rest of the paper, assigns a measure of importance to each node in each layer, accounting for the full interconnected structure of the multilayer network. However, in practical applications one is often interested in assigning a global measure of importance to each node, aggregating the information obtained from the different layers. The choice of the aggregation method is not trivial, it strongly influences the final estimation and might lead to wrong results.

\begin{figure*}[!t]
	\centering
	  \includegraphics[width=17cm]{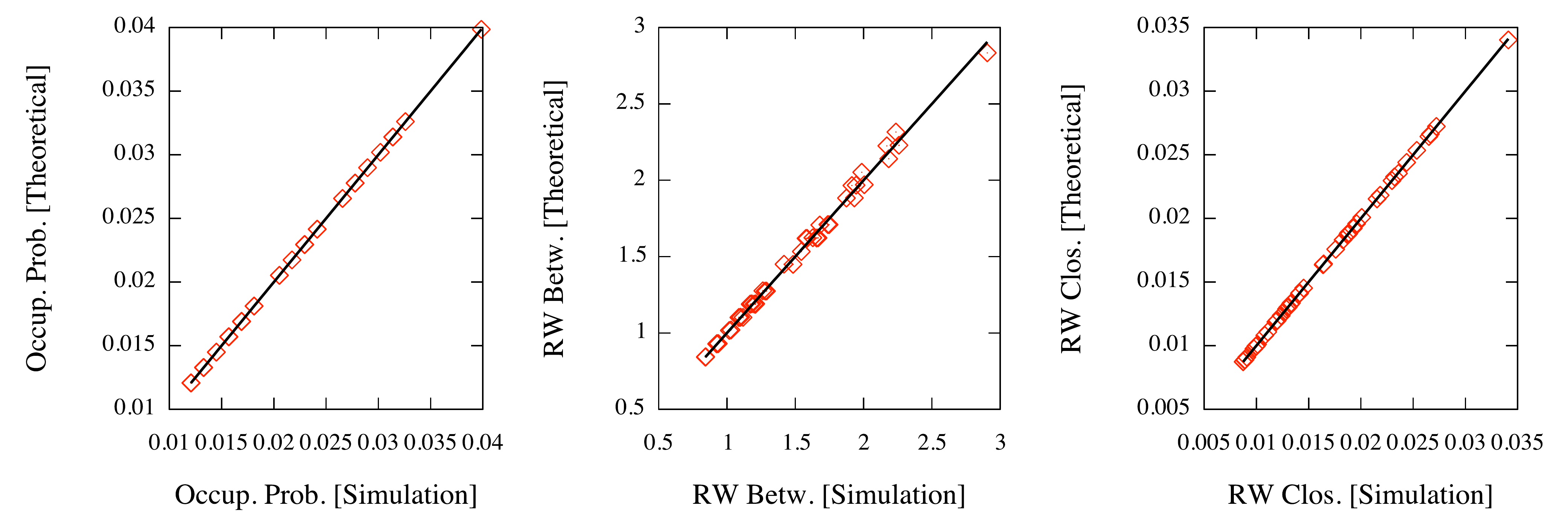}
	\caption{Representative example of the equivalence between the random walk centralities obtained from Monte Carlo simulations and the theoretical prediction. The case of a multiplex with two layers and 50 nodes is considered.}
        \label{fig:mplex-central-comparison}
\end{figure*}

However, this is not case for the framework discussed in the present study. In fact, the centrality $\Pi_{i\alpha}$ is calculated by inherently accounting for the interconnected structure of the whole system. We do not require to combine arbitrarily the information from different separate measures. In our framework, the most intuitive type of aggregation, i.e., summing up over layers, represents the unique and correct choice. Let $\pi_{i}=\Pi_{i\alpha}u^{\alpha}$ be the random walk centrality measure obtained by aggregating over layers. 
Here, $\pi_{i}$ indicates the probability of finding the walker in node $i$, regardless of the layer. It is worth noting that this probability is proportional to $s_{i\alpha}u^{\alpha}$, i.e., the multi-strength of node $i$. Therefore, in this specific case, the computation of the centrality by means of the aggregated network would provide the same result of the calculation accounting for the interconnected multiplex, if inter-layer edges are accounted for as self-loops. In the more specific case that the inter-layer edges have the same strength for all nodes, the random walk centrality will be just proportional to the degree of the aggregated network, without necessity of accounting for the self-loops. Unfortunately, this is no more the case for the centrality measures discussed in the rest of this study, where calculating the diagnostics from the aggregate might lead to wrong conclusions.

A ground-truth for this diagnostics can be obtained from numerical simulations of the random walk process in the multilayer network, where the larger the number of times the walker hits a node larger the random walk centrality of that node. In Fig.\,\ref{fig:mplex-central-comparison} we show the comparison between $\pi_{i}$ obtained from simulation and its theoretical prediction. As expected, the agreement is excellent and this equivalence holds regardless of the number of nodes in the network, the number of layers or their topology.

\vspace{0.25truecm}\emph{PageRank centrality.} We capitalize on this result to extend to interconnected networks a widely adopted measure of centrality, i.e., the PageRank \cite{brin1998}. A recent study in this direction has been reported in \cite{pagerank2013}, in the case of edge-colored graphs where the authors, exploiting the random walk interpretation of PageRank centrality, define the PageRank of a multiplex network by means of a random walk subjected to teleportation. In that study, the PageRank for nodes in the first layer is computed using the standard definition for a monoplex \cite{brin1998}, whereas the PageRank for nodes in the second layer is computed using the centrality information obtained from the first one. It is worth noting that this definition is limited to edge-colored graphs with only two layers, being any extension to a larger number of layers possible but very complicated from the mathematical point of view.

Here, we exploit the fact that PageRank centrality can be seen as the steady-state solution of the equation $p_{j}(t+1)=R^{i}_{j}p_{i}(t)$ in the case of monoplexes, where $R^{i}_{j}$ is the rank-2 transition tensor (i.e., the transition matrix) of a random walk where the walker jumps to a neighbor with rate $r$ and teleport to any other node in the network with another rate $r'$. For simplicity, we assume that $r'=1-r$ in the following. In the case of interconnected multilayer networks, the teleportation might occur to any other node in any layer. Depending on the application of interest, the walker can be teleported to other nodes with a rate that is specific to each layer. However, to keep the study as simple as possible, we consider the case with the same teleportation rate for all layers. Let $R^{i\alpha}_{j\beta}$ be the corresponding transition tensor, where the walker jumps to a neighbor with rate $r$ and teleport to any other node in the network with rate $1-r$. This rank-4 tensor is given by
\begin{equation}\label{RWPageRank}
	R^{i\alpha}_{j\beta} = rT^{i\alpha}_{j\beta} + \frac{(1-r)}{NL}u^{i\alpha}_{j\beta},
\end{equation}
where $u^{i\alpha}_{j\beta}$ is the rank-4 tensor with all components equal to 1.
The steady-state solution of the master equation corresponding to this transition tensor provides the PageRank centrality for interconnected multiplex networks. It is worth noting that the above definition is valid for all multiplexes where all nodes have out-going edges. If this is not the case, as in several real-world networks, Eq.\,(\ref{RWPageRank}) reduces to $R^{i\alpha}_{j\beta} = \frac{1}{NL}u^{i\alpha}_{j\beta}$ for all nodes $i$ with no out-going connections, ensuring the correct normalization of the transition tensor $R^{i\alpha}_{j\beta}$.

To compute the aggregate centrality of a node, accounting for the whole interconnected topology, we proceed as for the random walk occupation centrality previously discussed. Let $\Omega_{i\alpha}$ be the eigentensor of the transition tensor $R^{i\alpha}_{j\beta}$ (see Appendix\,\ref{sec:eigentensor} for details), denoting the steady-state probability to find the walker in node $i$ and layer $\alpha$. The multilayer PageRank is obtained by simply contracting the layer index of the eigentensor with the 1-vector: $\omega_{i} = \Omega_{i\alpha}u^{\alpha}$, i.e., by summing up over layers.

\vspace{0.25truecm}\emph{Random walk betweenness centrality.} The betweenness is a measure of network centrality which instead of accounting for topological centrality accounts for the importance of nodes in terms of dynamical processes that run over the network. In particular, the betweenness measures to which extent a node lies in the path between any two other nodes \cite{newman2005measure}. One can think of packets traveling in internet, in this case the betweenness measures the influence of nodes in the spreading of information. 

The most common betweenness is the shortest path betweenness \cite{Freeman77centralityin} where the centrality of a node $j$ is relative to the number of shortest paths, for any pair $(o,d)$ of \emph{origin} and \emph{destination} nodes, that pass through $j$. However, in real networks, entities (rumors, messages or packets over the Internet) that travel the network do not always take the shortest path \cite{Freeman91centralityin, Stephenson19891}. Consider, for instance, rumors that can be wandering around the network or packets trying to avoid overloaded routers. In such cases, the shortest path betweenness is not always a good proxy for the centrality of nodes. For these scenarios the random walk betweenness of a node $j$ is defined as the amount of random walks between any pair $(o,d)$ of nodes that pass through $j$ \cite{newman2005measure}. However, we discuss shortest-path betweenness centrality in interconnected multilayer networks at the end of the next section.

To analytically compute the number of random walks visiting a particular node, it is often convenient to use the concept of absorbing random walk, where the absorbing state is selected to be the destination node $d$ \cite{newman2005measure,Newman2010Book}. To extend this concept to the case of interconnected multilayer networks, we consider random walks that begin, end and pass by nodes in different layers while accounting for the existence of several replicas of the same node.

To extend the concept of random walks to interconnected networks, we define the absorbing transition tensor on a particular node $d$ by 
\begin{eqnarray} \label{eq:absTranTensor}
	\(T_{[d]}\)^{i\alpha}_{j\beta} = \left\{ 
		\begin{array}{l l}
		0 & \quad j=d\\
		T^{i\alpha}_{j\beta} & \quad j \ne d
	\end{array} \right.,
\end{eqnarray}
Random walkers governed by this transition tensor will vanish once they arrive to any absorbing state \cite{Newman2010Book}. Note that $T_{[d]}$ has one absorbing state for each replica of node $d$ in different layers.

It can be shown (see Appendix \ref{ap:timesOnNodeRWB}) that the average number of times a random walk (with origin in node $o$ in layer $\sigma$ and destination $[d]$) will pass by a node $j$ in layer $\beta$, regardless of the time step, is given by
\begin{eqnarray}
\({\tau_{[d]}}\)^{i\alpha}_{j\beta} = \[\(\delta - T_{[d]}\)^{-1}\]^{i\alpha}_{j\beta},
\end{eqnarray}
where $\delta^{i\alpha}_{j\beta}=\delta^{i}_{j}\delta^{\alpha}_{\beta}$. Note that the average number of times that the walk will visit node $j$ still depends on the layer where $j$ is located and on the originating layer $\sigma$. Since we are interested on node properties, regardless of the layer, we average over all possible starting layers $\sigma$ and aggregate the walks that pass through $j$ in the different layers, 
\begin{eqnarray}
	\(\tau_{[d]}\)_{j}^{o} &=& \frac{1}{L}\({\tau_{[d]}}\)^{o\sigma}_{j\beta}u^{\beta}u_{\sigma}.
\end{eqnarray}
The overall centrality vector is obtained by averaging over all possible origins and destinations:
\begin{eqnarray}\label{eq:anRWBetweenness}
\tau_{j}  = \frac{1}{N(N-1)}\sum\limits_{d = 1}^{N} \(\tau_{[d]}\)_{j}^{o}u_{o}.
\end{eqnarray}

The comparison between the values of $\tau_{i}$ obtained from simulation and theoretical prediction is shown in Fig.\,\ref{fig:mplex-central-comparison}. As expected, the results are in excellent agreement and it is worth remarking that the equivalence holds regardless of the number of nodes in the network, the number of layers or their topology.

It is worth investigating the influence of layer-to-layer correlation on random walk betweenness centrality. In fact, in general, a node on one layer can have different degree on other layers. However, there are situations where a node tends to be a hub in all layers, or it can be a hub in one layer with very low degree in another layer. To quantify the similarity between the degree of nodes across layers we make use of the Pearson coefficient, widely used to estimate the amount of linear degree-degree correlations and to assess the assortative/disassortative mixing patterns in single-layer networks \cite{newman2002assortative,newman2003mixing}. 

Analogously, we quantify the amount of linear positive/negative degree-degree correlation, or, equivalently, assortative/disassortative mixing, on different layers of an interconnected multiplex. Let $k_{i\alpha}$ indicate the degree of node $i$ in layer $\alpha$, and let $\kappa_{\alpha}$ indicate the average degree in the same layer. The inter-layer assortativity coefficient between $\alpha$ and another layer $\beta$ is given by 
\begin{eqnarray}
\mathcal{A}^{\alpha}_{\beta}=\frac{1}{\sigma_{[\alpha]}\sigma_{[\beta]}}\(k_{i\beta}-u_{i}\kappa_{\beta}\)\(k^{i\alpha}-u^{i}\kappa^{\alpha}\),
\end{eqnarray}
where 
\begin{eqnarray}
\sigma_{[\alpha]}&=&\sqrt{\(k_{i\alpha}-u_{i}\kappa_{\alpha}\)\(k^{i\alpha}-u^{i}\kappa^{\alpha}\)}\nonumber
\end{eqnarray}
is a scalar depending on layer $\alpha$ and $\sigma_{[\beta]}$ is a scalar depending on layer $\beta$. This coefficient is defined in the range between -1 (fully disassortative mixing) and +1 (fully assortative mixing). We define the two layers to show assortative (disassortative) inter-layer correlations if they are positively (negatively) correlated. Finally, if the degree of all vertices in layer $\alpha$ is not correlated to the degree of the same vertices in layer $\beta$, it is straightforward to verify that $\mathcal{A}^{\alpha}_{\beta}=0$. 

\begin{figure*}[!t]
	\centering
	  \includegraphics[width=18cm]{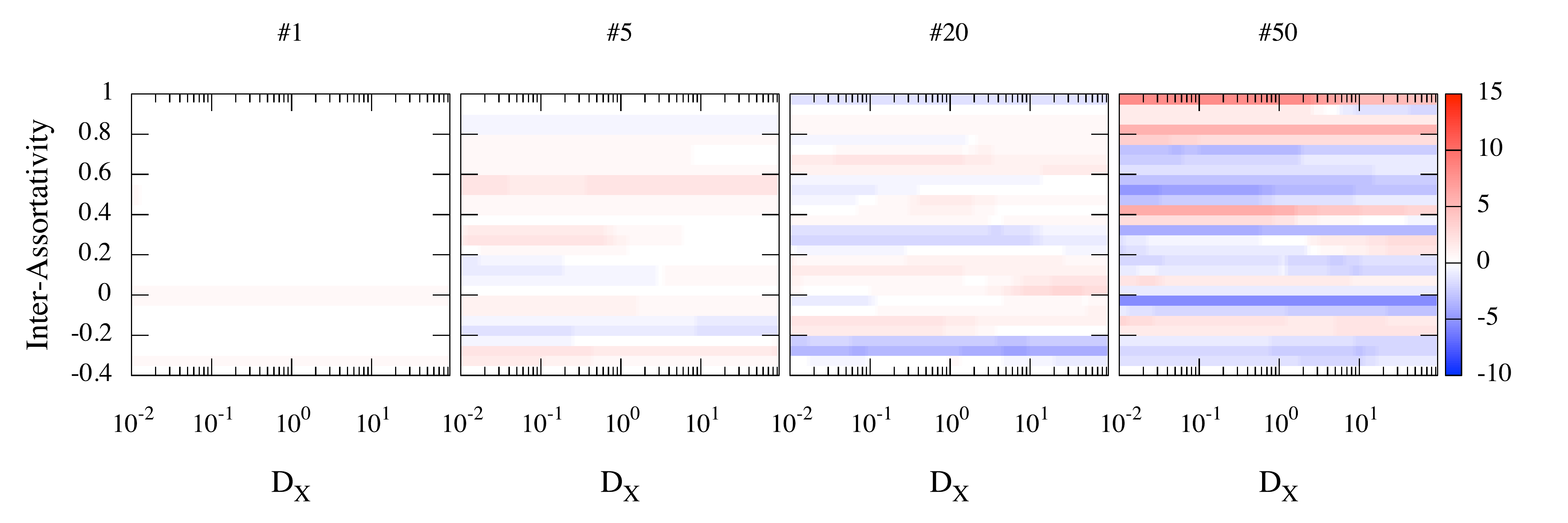}
	\caption{Random walk closeness centrality in interconnected multiplexes of two Barabasi-Albert networks. The relative difference between the rank in the multiplex and the rank in the corresponding aggregate is shown as a function of inter-layer connectivity and inter-layer assortativity. Results for four nodes whose centrality in the aggregated network (from left to right) is indicated on the top of each panel, are shown.}
        \label{fig:ranking-comparison-btw}
\end{figure*}

We build interconnected multiplexes of two Barabasi-Albert networks by varying the inter-layer correlations (see Appendix\,\ref{sec:assort-algorithm} for further details) and the intensity of the inter-layer link, the latter parameterized by $D_{X}$. For each configuration we calculate the random walk betweenness centrality of each node in the multiplex and in the corresponding aggregate. In Fig.\,\ref{fig:ranking-comparison-btw} is shown the relative difference between the ranks as a function of the inter-layer assortativity and $D_{X}$ for four representative nodes. It is evident that accounting for the whole interconnected structure alters the centrality of nodes changing their ranking.

\begin{figure*}[!t]
	\centering
	  \includegraphics[width=18cm]{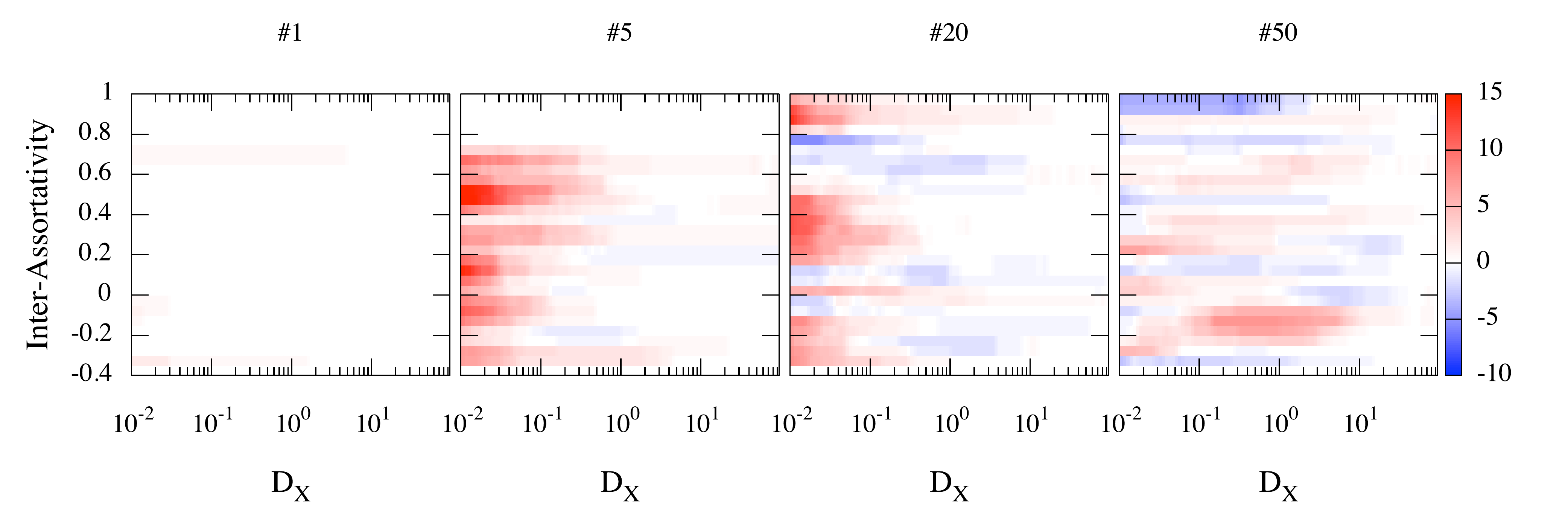}
	\caption{As in Fig.\,\ref{fig:ranking-comparison-btw} but for random walk closeness centrality.}
        \label{fig:ranking-comparison-cls}
\end{figure*}

\vspace{0.25truecm}\emph{Random walk closeness centrality.} The distance between two nodes in a network is given by shortest-path which separates them. The farness of an actor is given by the sum of all geodesics from that node to any other node. In general, the inverse of this farness provides a measure of the \emph{closeness} of the node. Such a diagnostic is related to how fast information is expected to spread from a given actor to the others in the network.

However, in many communication systems information does not spread along shortest-paths but it is more likely to follow random-walk-like paths. Note that we discuss shortest-path closeness centrality for interconnected multilayer networks at the end of the next section.

Therefore, a variant of the closeness for this type of systems is given by the random walk closeness centrality. In the case of monoplexes, it has been introduced to quantify how central a node is located regarding its potential to receive information randomly diffusing over the network \cite{noh2004random}.

In the case of interconnected multiplexes, we define the random walk closeness centrality of a node $i$ as the inverse of the average number of steps that a random walker, starting from any other node in the multilayer network, requires to hit $i$ for the first time. 

The average number of steps to reach a node $d$, starting from a node $s$, is known as mean first-passage time (MFPT) and it has been calculated exactly in the case of monoplexes by means of Kemeny-Snell fundamental matrix $Z$ \cite{lovasz1993random,zhang2011mean} of finite irreducible Markov chains \cite{kemeny1960finite} or by means of absorbing random walks \cite{kemeny1960finite,Newman2010Book}. In this study, we adopt the second approach as for the calculation of random walk betweenness centrality, where we use the transition tensor $T$ governing random walks over interconnected networks and the corresponding absorbing transition tensor $T_{[d]}$. The tensor
\begin{equation}
p_{j\beta}^{o\sigma}(t)= \(T^t_{[d]}\)_{j\beta}^{o\sigma}
\end{equation}
indicates the probability of visiting node $j$ in layer $\beta$, after $t$ time steps, considering that the walk originated in node $o$ in layer $\sigma$. This transition tensor is absorbing on node $d$ regardless of the layer and, consequently, any walker reaching an absorbing state will vanish, i.e., $p_{d\beta}^{o\sigma}(t)=0$ for any $\beta$ and $t$. The probability that the walker is absorbed in some node $d$ at a time $h$ equal or smaller than $t$, regardless of the layer, is given by
\begin{equation}
\(q_{[d]}\)^{o\sigma}(t)= u^{o\sigma}-\(T^t_{[d]}\)_{j\beta}^{o\sigma}u^{j\beta} \label{eq:probOfBeingAbsorbed}.
\end{equation}
Note that we have a rank-2 tensor $q$ for each choice of $d$ and we put in evidence this dependence by means of $[d]$. From each tensor $q$ we can calculate the probability that the first passage time for node $d$ is exactly $t$ by
\begin{eqnarray} 
\(q_{[d]}\)^{o\sigma}(h = t) &=& \(q_{[d]}\)^{o\sigma}(t)-\(q_{[d]}\)^{o\sigma}(t-1) \nonumber \\
&=& \[\(T^t_{[d]}\)-\(T^{t-1}_{[d]}\)\]_{j\beta}^{o\sigma}u^{j\beta}. \label{eq:pdf_closeness}
\end{eqnarray}
Considering the walk starting from node $o$ in layer $\sigma$, each tensor encoding the mean first passage time to node $d$, assumed to be the absorbing state, is obtained from Eq.\,(\ref{eq:pdf_closeness}) as
\begin{eqnarray} 
\(H_{[d]}\)^{o\sigma} = \sum\limits_{t=0}^{\infty} t \(q_{[d]}\)^{o\sigma}(h=t)  = \[\(\delta-T^{t}_{[d]}\)^{-1}\]_{j\beta}^{o\sigma}u^{j\beta}. \label{eq:meanTau}
\end{eqnarray}

The geometric series in Eq.\,(\ref{eq:meanTau}) converges since the maximum eigenvalue of $T_{[d]}$ is strictly smaller than one, and the sum can be calculated exploiting the self-similarity of the series. Note that the mean first passage time to $d$ still depends on the origin of the walk, i.e., node $o$ in layer $\sigma$. 

The average mean first passage time $h_{[d]}$ to node $d$ is obtained by averaging $\(H_{[d]}\)^{o\sigma}$ over all possible starting nodes and layers as
\begin{eqnarray} 
\label{eq:defMeanTime}
	h_{[d]} = \frac{1}{NL}u_{o\sigma}\(H_{[d]}\)^{o\sigma} + \frac{1}{N}\pi^{-1}_{[d]},
\end{eqnarray}
where $\pi_{[d]}$ is the occupation probability of node $d$ and the term $\frac{1}{N}\pi^{-1}_{[d]}$ is included explicitly to account for the average return time, that is not accounted for when using absorbing random walks. 

Finally, the random walk closeness centrality of node $d$ is defined by the inverse of $h_{[d]}$. We introduce the vector $\xi_{i}$ whose components are given by the inverse of the corresponding values of $h$.

The comparison between the values of $\xi_{i}$ obtained from simulation and theoretical prediction is shown in Fig.\,\ref{fig:mplex-central-comparison}. As expected, the results are in excellent agreement and it is worth remarking that the equivalence holds regardless of the number of nodes in the network, the number of layers or their topology.

As for the random walk betweenness centrality, we investigate the influence of layer-to-layer correlation on random walk closeness centrality. Following the same procedure, we build synthetic interconnected multiplexes by varying the inter-layer correlations and the intensity of the inter-layer link. In Fig.\,\ref{fig:ranking-comparison-cls} is shown the difference between the ranks as a function of the inter-layer assortativity and $D_{X}$ for four representative nodes. It is evident that accounting for the whole interconnected structure alters the centrality of nodes altering their ranking. We refer to the Supplemental Material for additional representative examples, both with synthetic and empirical networks, showing significant changes even in the top ranked nodes.


\subsection{Centrality based on topological properties}

\emph{Eigenvector centrality.} Among the numerous notions of centrality introduced to quantify the importance of nodes (and other components) in a network \cite{faust}, eigenvector centrality is among the oldest ones. A node $i$ has a high eigenvector centrality if its neighbors also have high eigenvector centrality, and the recursive nature of this notion yields a vector of centralities that satisfies an eigenvalue problem.

In the case of monoplexes, the \emph{eigenvector centrality vector}, whose components are the centralities of nodes according to \cite{bonacich1972,bonacich1972b}, is a solution of the tensorial equation $W^{i}_{j}v_{i}=\lambda_{1}v_{j}$, where $\lambda_{1}$ is the largest eigenvalue of $W^{i}_{j}$ and $v_{i}$ indicates the eigenvector centrality of node $i$.

A naive approach for the calculation of the importance of each node might be to project the interconnected topology to an aggregated monoplex, and to associate to each node the centrality he or she has in such an aggregated network. The main drawback of this approach is that it mixes the information from all layers with uncontrollable effects, as shown in the Supplemental Material for both synthetic and empirical networks.

Another attempt to extend this calculation to the case of multilayer networks might be to calculate the eigenvector centralities for each layer separately, to build the tensor $\bar{V}_{i\alpha}$ encoding the centrality of each node in each layer. The successive step would be to choose an heuristic aggregation of such centralities to assign a unique centrality measure to each node, regardless of the layer. However, the tensor $\bar{V}_{i\alpha}$ is not the solution of a unique eigenvalue problem but the combination of the solutions of $L$ different eigenvalue problems treated separately, therefore it is not the natural extension of the notion of eigenvector centrality to the realm of interconnected multilayer networks.

Instead, according to \cite{dedomenico2013mathematical}, this descriptor can be obtained as the solution of the tensorial equation
\begin{align}\label{eq:anEigenvector}
	M^{i\alpha}_{j\beta}\Theta_{i\alpha}=\lambda_{1} \Theta_{j\beta},
\end{align}
where $\lambda_{1}$ is the largest eigenvalue and $\Theta_{i\alpha}$ is the corresponding \emph{eigentensor} encoding the centrality of each node in each layer when accounting for the whole interconnected structure. The eigentensor can be obtained by means of an iterative procedure, as the power method in the case of monoplexes. A proof of the existence of such eigentensor is provided in Appendix\,\ref{sec:eigentensor}. Thus, the multilayer generalization of Bonacich's eigenvector centrality \cite{bonacich1972,bonacich1972b} is given by $\Theta_{j\beta} = \lambda_{1}^{-1}M^{i\alpha}_{j\beta}\Theta_{i\alpha}$ \cite{dedomenico2013mathematical}.

As already pointed out in the previous sections, the overall centrality of each node can be simply obtained by contracting over layers the centrality of each node in each layer, i.e., by $\theta_{i}=\Theta_{i\alpha}u^{\alpha}$.

At variance with the eigenvector centrality calculated from the monoplex aggregated \emph{before} the calculation and the one calculated by heuristically aggregating the centralities obtained separately, our measure is obtained from the mathematical extension of the original definition. The aggregation performed at the end of the calculation does not require any heuristic choice, because it is already accounting for the whole interconnected topology and, as we have previously shown, it is enough to contract the resulting eigentensor.

\vspace{0.25truecm}\emph{Katz centrality.} It is a well-known fact that eigenvector centrality can lead to wrong results in the case of directed networks. In fact, nodes with only outgoing edges have an eigenvector centrality of $0$ if Bonacich's definition is adopted. Moreover, in this case there are two leading eigenvectors, for in-going centrality and out-going centrality, requiring to distinguish between covariant and contravariant calculations. The Katz centrality \cite{katz1953new} attempts to solve the above problem by assigning a small amount $b$ of centrality to each node before calculating centrality. For monoplexes, the Katz centrality is given by $v_{j}=\((\delta-aW)^{-1}\)^{i}_{j}u_{i}$, where $a$ must be smaller than the largest eigenvalue and often one chooses $b = 1$.

Following a similar idea, we define the centrality tensor for each node in each layer as the solution of the tensorial equation
\begin{eqnarray}\label{eq:katz}
\Phi_{j\beta}=aM^{i\alpha}_{j\beta}\Phi_{i\alpha}+bu_{j\beta}, 
\end{eqnarray}
corresponding to the natural extension of the equation proposed by Katz to the case of interconnected multilayer networks. The solution is given by $\Phi_{j\beta}=\((\delta-aM)^{-1}\)^{i\alpha}_{j\beta}U_{i\alpha}$, where $\delta^{i\alpha}_{j\beta}=\delta^{i}_{j}\delta^{\alpha}_{\beta}$. As for the eigentensor centrality, this \emph{Katz centrality tensor} accounts for the whole interconnected topology and it is enough to contract it with the 1-vector to obtain the Katz centrality for each node, i.e., $\phi_{i}=\Phi_{i\alpha}u^{\alpha}$.


\vspace{0.25truecm}\emph{HITS centrality.} Similarly to the PageRank, another approach was introduced to rank Web sites with respect to their importance for users. This approach considers two different descriptors for each node, namely hub and authority \cite{kleinberg1999authoritative}. In fact, Web pages that point to an important page generally also point to other important pages, building a structure similar to a bipartite topology where relevant pages -- i.e., authorities -- are pointed by special Web pages -- i.e, hubs. It follows that nodes with high authority centrality are linked by nodes with high hub centrality while very influent hubs point to nodes which are very authoritative. Such a mechanism is described by two coupled equations which reduce to the two eigenvalue problems $\(WW^{\dag}\)^{i}_{j}v_{i}=\lambda_{1}v_{j}$ and $\(W^{\dag}W\)^{i}_{j}z_{i}=\lambda_{1}z_{j}$, where $W^{\dag}$ denotes the transpose of the adjacency tensor, $\lambda_{1}$ indicates the leading eigenvalue while $v_{i}$ and $z_{i}$ indicate hub and authority scores, respectively. The natural extension of the equations proposed by Kleinberg to the case of interconnected multilayer networks is given by
\begin{eqnarray}
\(M M^{\dag}\)^{i\alpha}_{j\beta} \Gamma_{i\alpha} &=& \lambda_{1} \Gamma_{j\beta},\\
\(M^{\dag} M\)^{i\alpha}_{j\beta} \Upsilon_{i\alpha} &=& \lambda_{1} \Upsilon_{j\beta},\\
\end{eqnarray}
where $\Gamma_{i\alpha}$ and $\Upsilon_{i\alpha}$ indicate hub and authority centrality, respectively. It is worth remarking that for undirected interconnected multiplexes, hub and authority scores are the same and equal to the corresponding eigenvector centrality. The hub and authority tensors should be contracted with the 1-vector to obtain the scores corresponding to each node regardless of the layer, i.e., $\gamma_{i} = \Gamma_{i\alpha}u^{\alpha}$ and $\upsilon_{i} = \Upsilon_{i\alpha}u^{\alpha}$, respectively.

\vspace{0.25truecm}\emph{Centrality measures based on shortest path.} For sake of completeness, in this paragraph we briefly extend centrality measures based on shortest paths, namely betweenness and closeness. 

Equivalently to the case of a monoplex, we define a path $\ell_{\[{o\sigma\rightarrow d\gamma}\]}\in\mathcal{P}_{\[{o\sigma\rightarrow d\gamma}\]}$, in the interconnected multilayer network, as an ordered sequence of nodes which starts from node $o$ in layer $\sigma$ and finishes in node $d$ in layer $\gamma$. We require that there exist an edge between any pair of consecutive nodes in $\ell$. Here, $\mathcal{P}_{\[{o\sigma\rightarrow d\gamma}\]}$ indicates the set of all possible paths between node $o$ in layer $\sigma$ and node $d$ in layer $\gamma$. For every path $\ell_{\[{o\sigma\rightarrow d\gamma}\]}$ it is possible to define a cost function $c\(\ell_{\[{o\sigma\rightarrow d\gamma}\]}\)$, usually depending on the weight of the edges the path traverses and on the application of interest, to account for the ``goodness'' of the path. Hence, the shortest path from node $o$ in layer $\sigma$ to node $d$ in layer $\gamma$ is the path
\begin{equation} \label{eq:shortestPathNodesLayer}
\ell^*_{\[{o\sigma\rightarrow d\gamma}\]} = \argmin\limits_{\ell'_{\[{o\sigma\rightarrow d\gamma}\]} \in \mathcal{P}_{\[{o\sigma\rightarrow d\gamma}\]}} c(\ell'_{\[{o\sigma\rightarrow d\gamma}\]}) 
\end{equation}
which minimizes the cost function. Using (\ref{eq:shortestPathNodesLayer}) we define the shortest path from node $o$ to node $d$, regardless of the layer, as
\begin{equation}
\ell^*_{\[{o\rightarrow d}\]} = \argmin\limits_{\sigma,\gamma\in \{1,2,\dots, L\}} \ell^*_{\[{o\sigma\rightarrow d\gamma}\]}.
\end{equation}
The centrality $\hat{\tau}_j$ of node $j$ is defined to be proportional to the number of times that node $j$, regardless of the layer, belongs to the set $\ell^*_{\[{o\rightarrow d}\]}$ for every possible origin-destination pair $(o,d)$.

The extension of the shortest-path betweenness centrality, defined in the case of monoplex networks in \cite{Freeman77centralityin}, is obtained by counting the number of shortest paths between any pair of \emph{origin} and \emph{destination} nodes $(o,d)$, that go though node $j$ in the interconnected structure. 

On the other hand, in the same spirit of monoplex networks, we define the shortest-path closeness centrality of a node $j$ in an interconnected multilayer topology as the average of the inverse of the cost of the shortest paths which start from any other node $o$ in the network. Thus, given the cost of a shortest path $c(\ell_{[o\rightarrow i]}^*)$ between node $i$ and node $o$, the shortest-path closeness centrality $\hat{\xi}_i$ can be easily computed by considering all possible origin nodes $o$.


\section{Conclusions and Discussion}\label{sec:conclusion}

We have presented the mathematical formulation of different measures of centrality in interconnected multilayer networks. We have grouped the definitions in two sets, those that are defined attending to a random navigation of the structure, and those defined from the topology itself. In the process, we have proven that, in general, these definitions differ from the naive addition of values between layers and adopt a complex nonlinear form. The results are ready to be used for complex networks analysis, and should be of interests in many interdisciplinary applications ranging from social sciences, to transportation networks.


\section*{Acknowledgements}

AA, MDD, SG, and AS were supported by the European Commission FET-Proactive project PLEXMATH (Grant No. 317614), the MULTIPLEX (grant 317532) and the Generalitat de Catalunya 2009-SGR-838. AA also acknowledges financial support from the ICREA Academia and the James S.\ McDonnell Foundation, and SG and AA were supported by FIS2012-38266. EO is supported by a PhD grant from the Region Ile-de-France.


\appendix


\section{Eigenvalue problem with tensors}\label{sec:eigentensor}

The eigenvalue problem for a rank-2 tensor, i.e., a standard matrix, is defined by $W^{i}_{j}v_{i}=\lambda v_{j}$. The extension of this problem to rank-4 tensors leads to the equation
\begin{eqnarray}
M^{i\alpha}_{j\beta}V_{i\alpha}=\lambda V_{j\beta}.
\end{eqnarray}
To solve this problem, it is worth noting that any tensor can be \emph{unfolded} to lower rank tensors \cite{kolda2009}. For instance, a rank-2 tensor like $W^{i}_{j}$, with $N^{2}$ components, can be flattened to a vector $w_{k}$ with $N^{2}$ components. In the case of the rank-4 multilayer adjacency tensor $M^{i\alpha}_{j\beta}$, although any unfolding is allowed, it is particularly useful for some applications to choose the ones flattening to a squared rank-2 tensor $\tilde{M}^{k}_{l}$ with $NL\times NL$ components, where $L$ indicates the number of layers \cite{gomez2013diffusion}. In fact, this unfolding produces as many block adjacency matrices, named \emph{supra-adjacency matrices} in some applications \cite{gomez2013diffusion,dedomenico2013random,cozzo2013cc}, as the number of permutations of diagonal blocks of size $N^{2}$, i.e., $L!$. However, such unfoldings do not alter the spectral properties of the resulting supra-matrix and can be used to solve the eigenvalue problem for rank-4 tensors. In fact, the solution of the eigenvalue problem
\begin{eqnarray}
\tilde{M}^{k}_{l}\tilde{v}_{k}=\tilde{\lambda}_{1}\tilde{v}_{l},
\end{eqnarray}
is a \emph{supra-vector} with $NL$ components which corresponds to the unfolding of the eigentensor $V_{i\alpha}$.


\section{Mean number of crossing times} \label{ap:timesOnNodeRWB}

Given $M$ random walks starting in node $o$ on layer $\sigma$ and ending when reaching node $d$, regardless of the layer, the expected number of times a random walk will pass by node $j$ on layer $\beta$ is given by
\begin{eqnarray}\label{eq:meanNumberOftimes}
\(\mathcal{T}_{[d]}\)^{o\sigma}_{j\beta} &=& \lim_{M \to \infty} \frac{1}{M} \sum\limits_{m=1}^{M} \sum\limits_{t=0}^{\infty} z_{j\beta}^{o\sigma}(t,m),
\end{eqnarray}
where $z_{j\beta}^{o\sigma}(t,m)=1$ if walk $m$ was visiting node $j$ in layer $\beta$ at time step $t$ and $z_{j\beta}^{o\sigma}(t,m)=0$ otherwise.

Following the frequentist interpretation, the probability of being in node $j$ in layer $\beta$ at time step $t$, provided that the walk originated in node $o$ in layer $\sigma$, is given by
\begin{eqnarray}\label{eq:probOfBeingAsLimit}
p_{j{\beta}}^{o{\sigma}}(t) = \lim_{M \to \infty} \frac{1}{M} \sum\limits_{m=1}^{m} z_{j\beta}^{o\sigma}(t,m).
\end{eqnarray}
Substituting (\ref{eq:probOfBeingAsLimit}) in (\ref{eq:meanNumberOftimes}) we obtain that 
\begin{eqnarray}
\(\tau_{[d]}\)^{o\sigma}_{j\beta} &=& \sum\limits_{t=0}^{\infty} p_{j{\beta}}^{o{\sigma}}(t) = \sum\limits_{t=0}^{\infty} \(T_{[d]}^{t}\)^{o\sigma}_{j\beta} \nonumber\\
&=& \[\(\delta - T_{[d]}\)^{-1}\]^{o\sigma}_{j\beta} 
\end{eqnarray}
where $T_{[d]}$ corresponds to the absorbing transition tensor defined in Eq.\,(\ref{eq:absTranTensor}). 


\section{Synthetic multiplex with given inter-layer assortativity}\label{sec:assort-algorithm}

In this appendix we explain the simple algorithm used to generate interconnected multiplex networks with a desired value $\mathcal{A}^{\star}$ of inter-layer assortativity. To keep the description as simple as possible, we adopt standard notation.

Let us start from a 2-layer multiplex with initial inter-layer assortativity $\mathcal{A}_{0}$. Inspired by the approach proposed by Xulvi-Brunet and Sokolov to modify assortative mixing in single-layer networks \cite{xulvi2004reshuffling}, we randomly choose two different vertices $i$ and $j$. The corresponding degrees on the two layers are $k_{i}^{1}$ and $k_{i}^{2}$ for vertex $i$, and $k_{j}^{1}$ and $k_{j}^{2}$ for vertex $j$. If $\mathcal{A}_{0}< (>) \mathcal{A}^{\star}$ we relabel node $i$ by $j$ in layer 2 if i) $|k_{i}^{1}-k_{i}^{2}|> (<)|k_{i}^{1}-k_{j}^{2}|$ \emph{and} ii) the new value $\mathcal{A}_{1}$ of the inter-layer assortativity of the resulting multiplex is such that $|\mathcal{A}_{1}-\mathcal{A}^{\star}|<|\mathcal{A}_{0}-\mathcal{A}^{\star}|$, otherwise we keep the network unmodified and we repeat the procedure. Although the convergence of this algorithm is not guaranteed, it has the advantage of not changing the single-layer global features of the second layer, as degree distribution, intra-layer assortativity, clustering and modularity. Moreover, our numerical experiments show that in the majority of considered cases the convergence to the desired value is obtained within a few iterations.


\begin{small}
\bibliographystyle{apsrev4-1} 

\bibliography{centrality}

\end{small}

\end{document}